\documentclass[aps,superscriptaddress,twocolumn,showpacs,preprintnumbers,amsmath,amssymb]{revtex4}
\usepackage{pst-all} 
\usepackage{multido}
\usepackage[dvips]{graphicx}
\usepackage{eucal} \usepackage{bm} 
\usepackage{float}
\usepackage{color}

\newcommand{\gsim}{\mbox{\raisebox{-0.6ex}{$\stackrel{>}{\sim}$}}\:}

\begin{document}
\title{
The eccentricity in heavy-ion collisions from Color Glass Condensate
initial conditions}

\medskip

\author{Azfar Adil}
\affiliation{
Physics Department, Columbia University,
538 West 120th Street, New York, NY 10027, USA
}
\author{Hans-Joachim Drescher}
\affiliation{
Frankfurt Institute for Advanced Studies (FIAS),
Johann Wolfgang Goethe-Universit\"at,
Max-von-Laue-Str.~1, 60438  Frankfurt am Main, Germany
}
\author{Adrian Dumitru}
\author{Arata Hayashigaki}
\author{Yasushi Nara}
\affiliation{
Institut f\"ur Theoretische Physik,
Johann Wolfgang Goethe-Universit\"at,
Max-von-Laue-Str. 1, 60438  Frankfurt am Main, Germany
}

\begin{abstract}
The eccentricity in coordinate-space at midrapidity of the overlap
zone in high-energy heavy-ion collisions predicted by the
$k_\perp$-factorization formalism is generically larger than expected
from scaling with the number of participants. We provide a simple
qualitative explanation of the effect which shows that it is not
caused predominantly by edge effects. We also show that it is quite
insensitive to ``details'' of the unintegrated gluon distribution
functions such as the presence of leading-twist shadowing and of an
extended geometric scaling window. The larger eccentricity increases
the azimuthal asymmetry of high transverse momentum
particles. Finally, we point out that the longitudinal structure of
the Color Glass Condensate initial condition for hydrodynamics away
from midrapidity is non-trivial but requires understanding of
large-$x$ effects.
\end{abstract}
\pacs{12.38.Mh,24.85.+p,25.75.Ld,25.75.-q}

\maketitle
\section{Introduction}
Elliptic flow~\cite{Ollitrault:1992bk} -- the azimuthal momentum
anisotropy of produced particles with respect to the reaction plane,
$v_2=\langle \cos(2\phi) \rangle$ -- is one of the key observables for
collective behavior in heavy ion collisions. The large $v_2$ observed
experimentally in semi-central Au+Au collisions at the Relativistic
Heavy-Ion Collider (RHIC)~\cite{RHICexperiments} is consistent with
non-viscous hydrodynamic expansion of a quark gluon plasma (QGP)
droplet~\cite{hydroreview}. On the other hand, a hadron ``plasma''
with small transverse pressure as initial condition underestimates
$v_2$ at midrapidity~\cite{hadronic}. These observations have
been interpreted as an indication for the formation of a QGP
with very small viscosity shortly after the instant of
collision.

Specific solutions of partial differential equations such as hydrodynamics
in general depend on the assumed boundary conditions.
Here, we focus on the initial conditions for the
subsequent evolution of the matter created in non-central Au+Au
collisions at RHIC.  Specifically, we consider the ``Color Glass
Condensate'' (CGC) initial conditions within a
$k_\perp$-factorization approach~\cite{KLN01}.

The elliptic flow $v_2$ in the final state is proportional to the
initial spatial eccentricity $\varepsilon$ in the plane
$\bm{r}_\perp=(r_x,r_y)$ perpendicular to the beam axis,
\begin{equation}
\label{eq:ecc}
\varepsilon = 
\frac{\langle r_y^{\,2}{-}r_x^{\,2}\rangle}
     {\langle r_y^{\,2}{+}r_x^{\,2}\rangle}~.
\end{equation}
The average is taken with respect to the initial energy density
distribution. The reaction plane is spanned by the beam ($z$) and
impact parameter ($x$) directions.

Ref.~\cite{Hirano:2005xf} observed that the CGC initial conditions
correspond to much larger eccentricity than the commonly assumed
scaling of the initial energy density with the local number of participants.
In fact, hydrodynamical simulations with such initial conditions were
shown to overestimate the measured $v_2$ even if dissipative effects in
the cool hadronic phase are taken into account (by using a hadronic
cascade~\cite{jam} rather than hydrodynamics). However,
ref.~\cite{Hirano:2005xf} argued that the different eccentricities
obtained from the CGC and Glauber model, respectively, arise from
mainly the low-density edges of the overlap region where the CGC
approach is less trustworthy.

In this paper we show that more sophisticated models for the nuclear
unintegrated gluon distribution $\phi(x,k_\perp)$ than those employed
in the original KLN approach~\cite{KLN01} and in~\cite{Hirano:2005xf}
also predict much larger eccentricity than the Glauber approach. In
particular, we shall calculate $\varepsilon$ using gluon distributions
which account not only for saturation but also for extended geometric
scaling (anomalous dimension) and leading-twist shadowing, both of
which are important for transverse momenta above the saturation
momentum $Q_s$. We also provide a simple qualitative explanation of
the large eccentricity in the CGC approach which motivates that it
does not originate from pure ``edge effects''. We therefore expect
that both the large eccentricity of the transverse overlap region as
well as the in-plane ``twist'' of produced gluons~\cite{Adil:2005bb}
are rather generic predictions of the $k_\perp$-factorization
approach. Finally, using a simple model for energy loss induced by the
dense medium, we show how high-$p_\perp$ partons could probe the
initial distribution of produced gluons in coordinate space.

\section{Gluon production}

In the $k_\perp$-factorization approach
the number distribution of produced gluons is given by
\begin{eqnarray}
  \frac{dN_g}{d^2 r_{\perp}dy}&=&
  \frac{4N_c}{N_c^2-1} \int^{p_\perp^\mathrm{max}}\frac{d^2p_\perp}{p^2_\perp}
  \int d^2k_\perp \;\alpha_s\,
    \phi_A(x_1,\bm{p}_\perp^2)
  \nonumber\\
  && {} \times      \phi_B(x_2,(\bm{p}_\perp{-}\bm{k}_\perp)^2)~,
 \label{eq:ktfac}
\end{eqnarray}
with $N_c=3$ the number of colors.  Here, $p_\perp$ and $y$ denote the
transverse momentum and the rapidity of the produced gluons,
respectively. The light-cone momentum fractions of the colliding gluon
ladders are then given by $x_{1,2} = p_\perp\exp(\pm y)/\sqrt{s_{NN}}$,
where $\sqrt{s_{NN}}$ denotes the center of mass energy.  We set
$p_\perp^\mathrm{max}$ such that the minimal saturation scale
$Q_s^\mathrm{min}(x_i)$ in the above integration is
$\Lambda_{QCD}=0.2$ GeV.  We note that the eccentricity {\em at
midrapidity} is rather insensitive to the choice of
$p_\perp^\mathrm{max}$~\cite{Kuhlman:2006qp}.  The KLN
approach~\cite{KLN01} employs the following unintegrated gluon
distribution function at $x\ll1$:
\begin{equation}
\label{eq:uninteg}
  \phi(x,k_\perp^2;\bm{r}_\perp)\sim
  \frac{1}{\alpha_s(Q^2_s)}\frac{Q_s^2}
   {{\rm max}(Q_s^2,k_\perp^2)}~.
\end{equation}
$Q_s$ denotes the saturation momentum at the given momentum fraction
$x$ and transverse position $\bm{r}_\perp$. The overall normalization
of $\phi$ is determined by the multiplicity at midrapidity for the
most central collisions. In the KLN approach, the saturation scale for
a nucleus in $A{+}B$ collisions is then parameterized as
\begin{equation}
 Q^2_{s\;A,B}(x,\bm{r}_\perp) = 
   2\,{\rm GeV}^2\left(\frac{n_\mathrm{part}^{A,B}(\bm{r}_\perp)}{1.53}\right)
                       \left(\frac{0.01}{x}\right)^\lambda,
		       \label{eq:qs}
\end{equation}
where $n_\mathrm{part}(\bm{r}_\perp)$ is the transverse density of
participants. It is calculated from the thickness functions $T_A$ and $T_B$:
\begin{eqnarray}
 n^A_\mathrm{part}(\bm{r}_\perp) &= &
   T_A(\bm{r}_\perp + \bm{b}/2) \nonumber\\
 &\times& (1-(1-\sigma_{NN}T_B(\bm{r}_\perp - \bm{b}/2)/B)^B)~.
 \label{eq:npart}
\end{eqnarray}
$\bm{b}=(b,0)$ is the impact parameter vector in the transverse
plane. The nucleon-nucleon inelastic cross section at
$\sqrt{s_{NN}}=200$~GeV is $\sigma_{NN}=42$~mb.

The form $Q_s^2(x)\sim x^{-\lambda}$ with $\lambda\approx0.2-0.3$ is
motivated by deep inelastic scattering (DIS) data from the Hadron
Electron Ring Accelerator (HERA) for $x<0.01$~\cite{GBW}. The
parameterization~(\ref{eq:qs}) leads to an average saturation momentum
at midrapidity of about 1.5~GeV for central Au+Au collisions at
$\sqrt{s_{NN}}=200$~GeV, which is in line with standard
estimates~\cite{KLN01}.

The KLN model~(\ref{eq:uninteg}) exhibits a transition from
saturation, $\phi(k_\perp^2)\simeq{\rm const.}$, directly to the DGLAP
regime $\phi(k_\perp^2)\sim 1/k_\perp^2$ at $k_\perp=Q_s$; it does not
incorporate leading-twist shadowing since $\phi(k_\perp>Q_s)$ is
proportional to the density of participants at a given transverse
coordinate. On the other hand, it is well known that DIS data from
HERA exhibit approximate ``geometric scaling'' even for
$Q^2>Q_s^2$~\cite{Stasto:2000er}. This indicates that at small $x$ the
anomalous dimension $\gamma(k_\perp)$ of the gluon distribution
function evolves smoothly towards its DGLAP limit of $\gamma=1-{\cal
O}(\alpha_s)\simeq1$ as $k_\perp/Q_s$ increases. This is a consequence
of quantum evolution of $\phi$ with $\log\;
1/x$~\cite{ScalingViol}. The unintegrated gluon distribution can then
be written as
\begin{equation}
\label{eq:unintegx}
  \phi(x,k^2_\perp;\bm{r}_\perp)
   \sim \frac{1}{\alpha_s(Q^2_s)}\;{\rm min}\left(1,
\left(\frac{Q^2_s}{k^2_\perp}\right)^{\gamma(x,k_\perp^2)}\right)~.
\end{equation}
Due to $0<\gamma< 1$, $\phi(k_\perp^2>Q_s^2)$ from~(\ref{eq:unintegx})
is harder than the one from the original KLN model. The contribution
from the extended geometric scaling regime to the $p_\perp$-integrated
multiplicity is small when $Q_s$ is large. Since the eccentricity
$\varepsilon$ involves an integration over a wide range of saturation
momenta, however, it appears less obvious whether or not the geometric
scaling window above $Q_s$ plays a role. Here, we employ the
following parameterization of $\gamma$ for $k_\perp>Q_s$~\cite{DHJ}:
\begin{equation}
\label{eq:anom}
 \gamma(x_i,k_{T}^2) = \gamma_s + (1-\gamma_s)
  \frac{\log(k_\perp^2/Q_{s,i}^2)}
    {\lambda Y_i + 1.2\sqrt{Y_i} + \log(k_\perp^2/Q_{s,i}^2)}~,
\end{equation}
where $\gamma_s=0.627$ is the anomalous dimension corresponding to
leading-order (LO) BFKL evolution with saturation boundary conditions
in the saddle point approximation~\cite{ScalingViol}. The rapidities
$Y_i$ of the fusing gluon ladders are related to their momentum
fractions $x_i$ by $Y_i=\log(1/x_i)$.

The extended geometric scaling window corresponds to the region of
$k_\perp$ where $\gamma$ evolves from $\gamma_s$ to $\sim1$. For
asymptotically small $x$, one can neglect the term $\sim\sqrt{Y}$
in~(\ref{eq:anom}) and the upper end of the extended geometric scaling
window is then given by $Q_{gs}(x)\sim
Q_s^2(x)/Q_0$~\cite{ScalingViol}; for the central region of RHIC
collisions, the approach to the DGLAP regime is mainly driven by the
subasymptotic $\sim\sqrt{Y}$ term. The above form of $\gamma$
reproduces the evolution of inclusive $p_\perp$-distributions from
minimum-bias $d+Au$ collisions with rapidity rather well~\cite{DHJ}.

Finally, we shall also compute $\varepsilon$ using an unintegrated
gluon distribution obtained from an {\em Ansatz} for the dipole
forward scattering amplitude
$N(x,\bm{u}_\perp)$~\cite{KKT03,Braun:2000bh}:
\begin{equation}
\label{eq:uninteg2}
  \phi(x,k^2_\perp;\bm{r}_\perp) \sim \frac{1}{\alpha_s}\int
d^2\bm{u}_\perp e^{-i\bm{k}_\perp\cdot\bm{u}_\perp} \nabla^2_\perp
N(x,\bm{u}_\perp),
\end{equation}
where $\bm{u}_\perp$ is the transverse size of the dipole,
conjugate to $\bm{k}_\perp$. A viable model for
the unitarized dipole profile is~\cite{KKT03,DHJ}
\begin{equation}
  N(x,\bm{u}_\perp) = 1 - \exp\left[
       -\frac{1}{4}(u_\perp^2Q_s^2)^{\gamma(x,u_\perp^2)}\right]~.
\label{UnitaryDipole}
\end{equation}
where the anomalous dimension $\gamma(x,u_\perp^2)$ is of the same
form as (\ref{eq:anom}) but with the replacement $k_\perp^2\to1/u_\perp^2$.
This dipole profile describes the transitions from saturation
($N(u_\perp)\sim1$) to the leading-twist extended geometric
scaling ($N(u_\perp)\sim(u^2_\perp Q^2_s)^\gamma$) 
and DGLAP regimes ($N(u_\perp)\sim u^2_\perp Q^2_s$). Leading-twist
shadowing is present in the extended geometric scaling window since
with $Q_s^2\sim A^{1/3}$, $N\sim A^{\gamma/3}$
exhibits a weaker dependence on the nuclear thickness.

\section{Results}

\begin{figure}[tb]
\begin{center}
\includegraphics[width=8.5cm,clip]{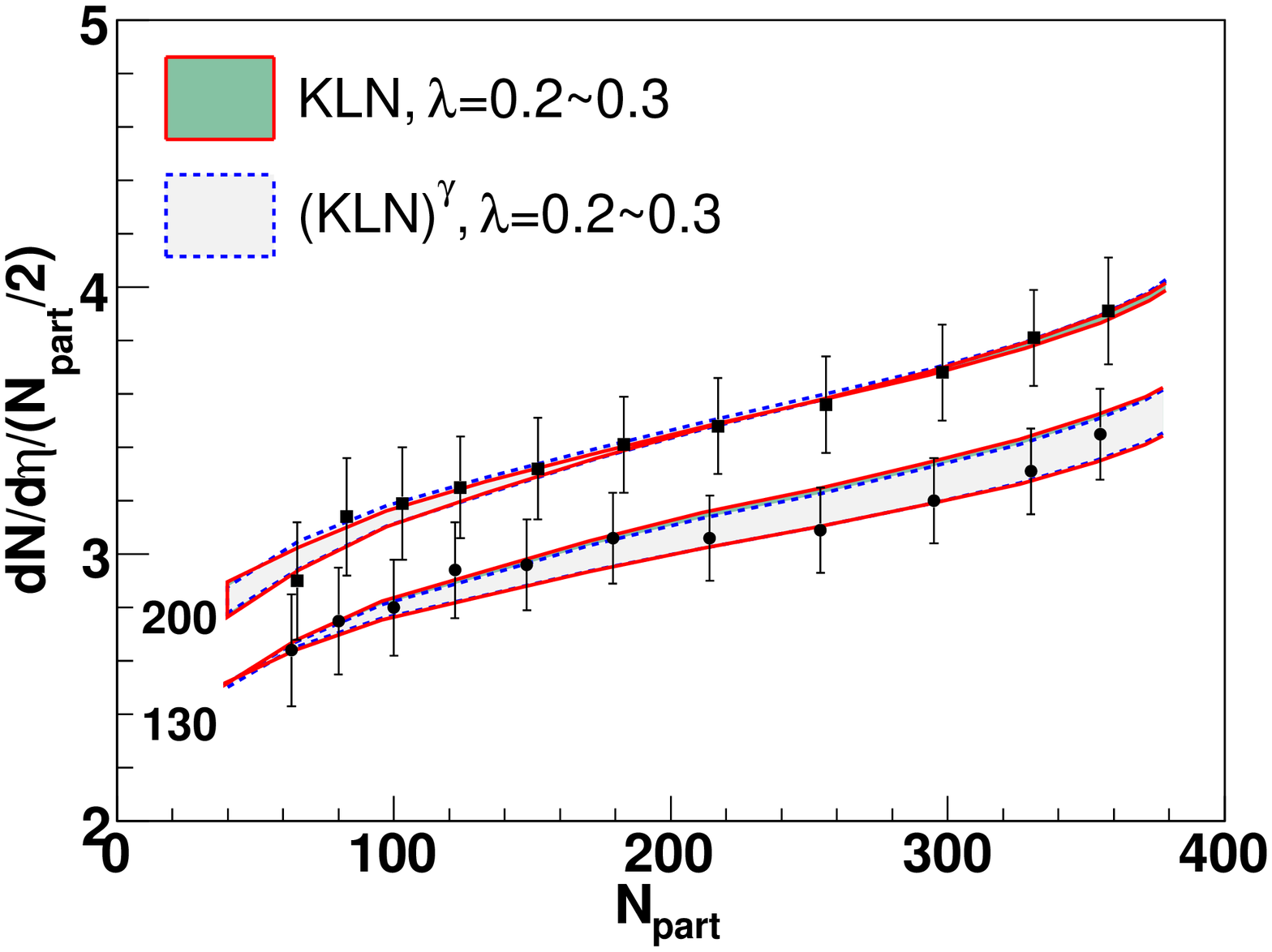}
\includegraphics[width=8.5cm,clip]{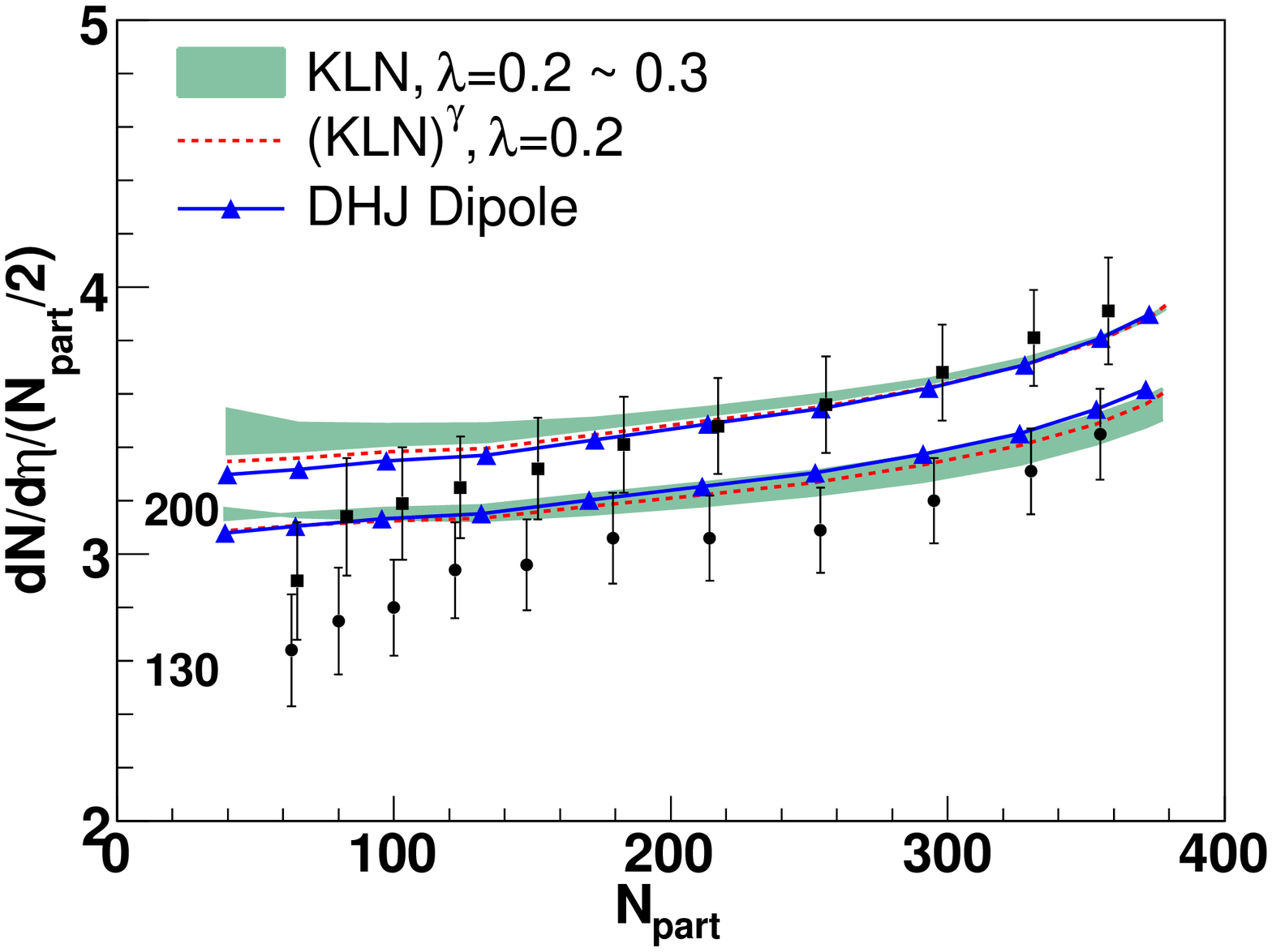}
\caption{(Color online) Centrality dependence of the multiplicity
in Au+Au collisions as compared to PHOBOS data~\cite{phobos} for
$\sqrt{s_{NN}}=130$ and 200~GeV.  Running (fixed) coupling is used in the
upper (lower) panel. $\lambda$ denotes the growth rate of the
saturation momentum with $\log\,1/x$. The curves correspond to various
parameterizations of the unintegrated gluon distribution function, see
text for detailed explanations.}
\label{fig:dndy}
\end{center}
\end{figure} 

In this section, we show the results of a numerical integration of
eq.~(\ref{eq:ktfac}), using the above-mentioned models for the
unintegrated gluon distribution $\phi(x,k_\perp)$. We shall employ
both fixed ($\alpha_s=0.5$) and LO running coupling constants,
$\alpha_s(Q^2_s)=4\pi/9\log(Q_s^2/\Lambda_{QCD}^2)$~\cite{fn:cut}.

We first check the centrality dependence of the multiplicity.  In
Fig.~\ref{fig:dndy} we compare the calculated number of gluons to
PHOBOS data~\cite{phobos} for charged hadrons, assuming that the
hadron yield is proportional to the yield of produced gluons.
``(KLN)'' denotes the original KLN parameterization of the
unintegrated gluon distribution function~(\ref{eq:uninteg}),
``(KLN)$^\gamma$'' its extension from eq.~(\ref{eq:unintegx}) to
include the smooth transition of the anomalous dimension and
leading-twist shadowing, and ``DHJ Dipole'' the unitarized dipole
profile from eqs.~(\ref{eq:uninteg2},\ref{UnitaryDipole}).  We find
that all models are consistent with the data within error bars.  As a
side-remark, we note that the explicit numerical integration of
eq.~(\ref{eq:ktfac}) over $\bm{r}_\perp$ improves the KLN results at
midrapidity, especially in the high $N_\mathrm{part}$ region, as
compared to the $Q_s^2(\bm{r}_\perp)\to\langle Q_s^2\rangle$
approximation employed in Refs.~\cite{KLN01}.  We refer to
Ref.~\cite{Kuhlman:2006qp} for a nice explanation of the positive
slope of the curve; also, to ref.~\cite{Krasnitz:2002mn} for a
computation of the centrality dependence of the multiplicity within
the classical McLerran-Venugopalan model and its relation to the KLN
approach.

\begin{figure}[t]
\begin{center}
\includegraphics[width=9cm,clip]{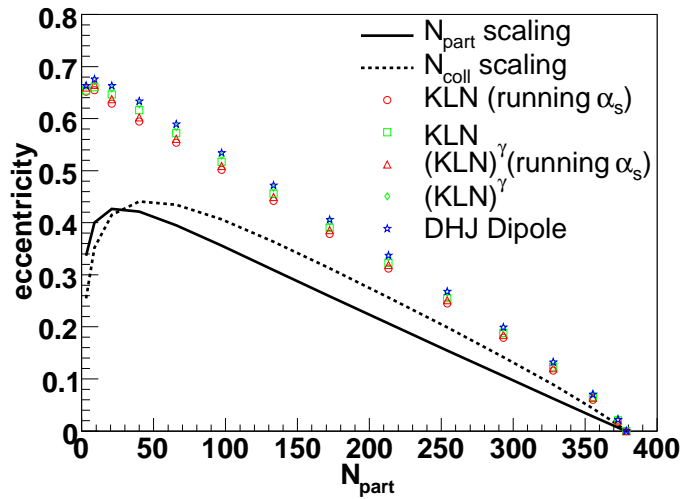}
\caption{(Color online) Initial spatial eccentricity $\varepsilon$ at
midrapidity as a function of the number of participants for
200\,$A$\,GeV Au+Au collisions from various models.  For comparison,
we also show initial conditions where the initial parton density at
midrapidity scales with the transverse density of wounded nucleons
(full line) and of binary collisions (dotted
line)~\cite{Kolb:2001qz}.}
\label{fig:ecc}
\end{center}
\end{figure}
Next, in Fig.~\ref{fig:ecc}, we compare the eccentricity for the
above-mentioned unintegrated gluon distribution functions.  Regardless
of running or fixed coupling, all saturation models yield almost the
same $\varepsilon$ as a function of centrality while that predicted by
Glauber-type models is much lower. The latter is commonly used as
initial condition for hydrodynamical simulations~\cite{Kolb:2001qz},
assuming that the energy density in the transverse plane scales with
either the number of participants or the number of collisions
(dominance of hard processes), respectively. The fact that
all of the above models for $\phi$ predict about the same $\varepsilon$
already indicates that the large anisotropy in coordinate space might
be generic to the $k_\perp$-factorization formula~(\ref{eq:ktfac}).

\begin{figure}[tb]
\begin{center}
\includegraphics[width=5cm,clip]{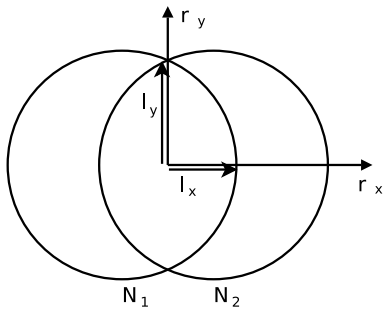}
\caption{Along the line $\vec{l}_x$, the density of gluons 
in the CGC case falls off more rapidly than in Glauber type
models. Along $\vec{l}_y$, the collision is symmetric and the CGC
gluon density behaves similar to the Glauber model. See text for explanation.}
\label{fig:Almond}
\end{center}
\end{figure} 
The large predicted eccentricity could be understood
qualitatively in the following way. In the $k_\perp$-factorization
approach the number of produced gluons scales approximately with the
{\em smaller} of the saturation scales of the two nuclei~\cite{KLN01}.
This is valid not only as a function of rapidity but also as a
function of the transverse coordinate~\cite{EtQsmax}:
\begin{equation}
\frac{dN}{d^2\bm{r}_\perp dy}
   \sim \min(Q^2_{s,1}(y,\bm{r}_\perp),Q^2_{s,2}(y,\bm{r}_\perp)).
\end{equation}
To understand this, approximate the local
unintegrated gluon distribution by $\phi_A(k_\perp)
\simeq\Theta(Q_{s,{\rm min}}-k_\perp)$. Scaling with $Q_{s,{\rm min}}$
follows also from analytical solutions of the classical Yang-Mills
equations for asymmetric collisions~\cite{DMcL}.

Fig.~\ref{fig:Almond} depicts two paths
away from the center of the overlap region. Along $\vec{l}_x$
the number density scales, to logarithmic accuracy, as
\begin{equation}
\label{eq:scale1}
\rho_{\mathrm{CGC}}(r_x,0) \sim Q^2_{s,1}(r_x,0) 
\sim n_{\mathrm{part},1}(r_x,0) ~.
\end{equation}
On the other hand, in the opposite direction along $-\vec{l}_x$ we have
\begin{equation}
\label{eq:scale1b}
\rho_{\mathrm{CGC}}(r_x,0) \sim Q^2_{s,2}(r_x,0) 
\sim n_{\mathrm{part},2}(r_x,0) ~.
\end{equation}
In the wounded nucleon model, the density scales
with the {\em total} number of participants from both nuclei,
\begin{equation}
\label{eq:scale2}
\rho_{\mathrm{Glauber}}(r_x,0)  \sim n_{\mathrm{part},1}(r_x,0)
+n_{\mathrm{part},2}(r_x,0)~,
\end{equation}
which of course drops less rapidly with $r_x$ since the drop of the
density towards the edge of the more dilute nucleus is at least partly
compensated by the increasing number of participants from the other
nucleus.

Along the line $\pm\vec{l}_y$, in turn, both saturation scales are
equal and so
\begin{eqnarray}
\rho_{\mathrm{CGC}}(0,r_y) &\sim&
 (n_{\mathrm{part},1}(0,r_y)+n_{\mathrm{part},2}(0,r_y))/2 \nonumber\\ 
 &\sim& \rho_{\mathrm{Glauber}}(0,r_y) ~.
\end{eqnarray}
Since the number density along $l_x$ drops more rapidly for the CGC
initial conditions than for the Glauber model, while both behave the
same along $l_y$, the resulting eccentricity is higher for the former.
Thus, the large eccentricity is a feature inherent to the
$k_\perp$-factorization formula. 

Contrary to simple geometrical overlap models, the CGC also predicts a
non-trivial longitudinal structure of the initial condition. Even
under the simplest assumption of {\em local} quantum evolution, which
implies that the $\bm{r}_\perp$-dependence of $\phi_{A,B}$ arises
exclusively via $Q_{s\;A,B}^2(x,\bm{r}_\perp)\sim n_{\rm part}^{A,B}
(\bm{r}_\perp)$ at any
rapidity, moments of the density of produced gluons in coordinate
space do depend on $y$~! For example, the ``center of gravity'' $\langle
r_x\rangle$ does not vanish at $y\neq0$, as
already discussed in~\cite{Adil:2005bb}. 

\begin{figure}[tb]
\begin{center}
\includegraphics[width=7cm,clip]{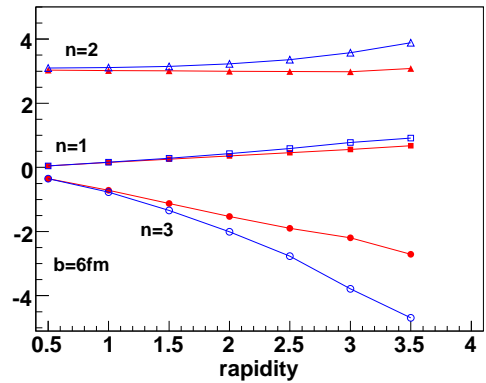}
\caption{(Color online) Moments ${\langle(r_x-\langle
  r_x\rangle)^n\rangle}$ (in fm$^n$) of the coordinate-space
  distribution of produced gluons as a function of rapidity. Open
  (closed) symbols correspond to the {\em Ansatz}~(\ref{eq:uninteg})
  with (without) the additional factor of $(1-x)^4$. The $n=1,3$
  moments have been scaled up by a factor of 10.}
\label{fig:x-moments}
\end{center}
\end{figure} 
For gluon production at non-zero rapidity the momentum fraction $x$ of
one of the fusing gluon ladders grows like $\sim\exp(y)$. In the KLN
approach, large-$x$ effects are incorporated into the unintegrated
gluon distribution by the substitution $\phi(x,k_\perp^2)\to
\phi(x,k_\perp^2) (1-x)^4$. This is, of course, a rather
qualitative model of large-$x$ effects which does not treat their
$Q^2$ dependence etc. 

Fig.~\ref{fig:x-moments} shows the evolution of various moments of the
energy density of produced gluons with rapidity. The longitudinal
structure is clearly non-trivial since odd moments of $r_x$ no longer
vanish at $y\neq0$. Hydrodynamical solutions will exhibit a
``left-right'' asymmetry (i.e.\ $v_1\neq0$) due to the fact that
$\langle(r_x-\langle r_x\rangle)^3\rangle\neq0$~\cite{brachi}. Hence,
as compared to the geometry at $y=0$ illustrated in
Fig.~\ref{fig:Almond}, not only does the center of the ``overlap''
shift with rapidity but the entire shape
changes. Fig.~\ref{fig:x-moments} also shows that the $(1-x)^4$
factor, which models large-$x$ effects, strongly affects the third
moment of $r_x$ at $y\gsim2$. (Other ``details'' such as $p_\perp^{\rm
max}$ from eq.~(\ref{eq:ktfac}) are also much more important at large
$y$, which is another reason why there the numbers from
Fig.~\ref{fig:x-moments} can not be trusted quantitatively.)  This
result indicates that our present limited understanding of ``large-$x$''
effects within the CGC formalism inhibits more quantitative
calculations of the initial conditions for hydrodynamic models over a
broad range of $y$~\cite{LargeX}. In order to compute the observable
momentum-space distributions of particles from these initial
conditions one needs to solve for the subsequent thermalization and
three-dimensional hydrodynamical expansion~\cite{HN} of produced
gluons, which is beyond the scope of the present note.

However, high-$p_\perp$ particles may probe the initial density
distribution via final-state interactions such as energy loss.
We employ a simple model for radiative energy loss of fast partons to
illustrate the qualitative effect.

In a dense QCD medium the induced radiative energy loss reduces the
initial transverse momentum $p_{\perp}^{0}$ of a produced hard parton
prior to its fragmentation into hadrons.  The ``quenching factor''
$(1-\mu\chi)$ depends in general on the initial parton rapidity $y$,
its initial transverse momentum $p_{\perp}^0$ as well as on the local
density~\cite{Gyulassy:2003mc}.  Here, we consider only mid-rapidity
$(y=0)$ partons and further simplify the calculation by neglecting the
mild $p_\perp$ dependence of the fractional energy loss.  We use the
model proposed in~\cite{Drees:2003zh} and further developed
in~\cite{Horowitz}. $\mu$ is a parameter set to reproduce the observed
average jet quenching at $b=0$. For eikonal (straight-line) jet
trajectories, the opacity $\chi$ of the medium is a function of the
initial production point $\bm{r}_\perp$ of the jet and of its
azimuthal direction of propagation $\hat{\bm{v}}_\perp =
(\cos\varphi,\sin\varphi)$,
\begin{eqnarray} \label{opac}
\chi(\bm{r}_\perp,\hat{\bm{v}}_\perp;b)&=& l_0 \int_{l_0}^\infty dl \, 
  \frac{l-l_{0}}{l} \,
  \rho(\bm{r}_\perp+l\hat{\bm{v}}_\perp;b)~.
\end{eqnarray}
where $\rho(\bm{r}_\perp,b)$ denotes the local bulk density.
Here, we assume a formation time of the medium of
$l_{0}=0.2$~fm. Larger formation times tend to increase the final
azimuthal asymmetry~\cite{pantuev}. Eq.~(\ref{opac}) accounts
for non-dissipative one-dimensional expansion of the medium along the
beam axis.

For simplicity, we approximate the initial $p_\perp$-distribution of
hard partons with $p_\perp^0>Q_{gs}$ by a power law
${d\sigma}/{d^{2}p_\perp^0 dy}\sim{(p^0_{\perp})^{-n(p^0_{\perp})}}$,
where $n(p^0_{\perp})$ is a slowly varying function which we calculate
in the DGLAP leading-twist approximation. This is justified since, by
definition, the anomalous dimension $\gamma\to\gamma_{\rm DGLAP}$ in
the high-$p_\perp$ regime.  The final transverse momentum of a given
parton then is $p_\perp=p_{\perp}^0{(1-\mu\chi)}$. We define the
quantity $R_{AA}(p_\perp,\varphi;b)$ as the ratio of the initial to
the final transverse momentum distributions, omitting fragmentation
into hadrons:
\begin{eqnarray}
R_{AA}(p_\perp,\varphi;b) &=& \Big<
f_g(p_\perp)(1-\mu\chi)^{n_g(p_\perp)-2}\nonumber\\
 & & \hspace{-1cm} + \; (1-f_g(p_\perp))
(1-\frac{4}{9} \mu\chi)^{n_q(p_\perp)-2}\Big>~,
\label{raaunav}
\end{eqnarray}
where $f_{g}(p_{\perp})$ is the fraction of gluons and
$n_{q/g}(p_{\perp})$ are the quark and gluon power laws,
respectively~\cite{commentRaa}. The color factor $C_F/C_A=(N^2-1)/(2N^2)=4/9$
accounts for the weaker coupling of quark jets to the medium. The
average in~(\ref{raaunav}) is performed over the initial production points
$\bm{r}_\perp$ of the jets. In the DGLAP regime, these are distributed
according to the local number of binary collisions, $\sim
T_A(\bm{r}_\perp+\bm{b}/2)\, T_B(\bm{r}_\perp-\bm{b}/2)$.

\begin{figure}[tb]
\begin{center}
\includegraphics[width=9cm,clip]{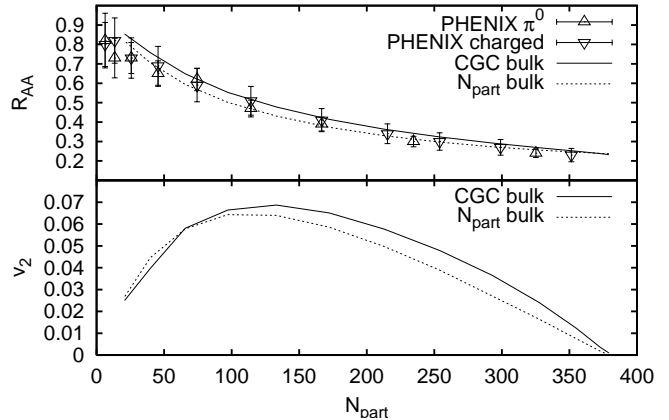}
\caption{Top: suppression factor $R_{AA}(p_\perp;b)$ at
  $p_\perp\simeq10$~GeV as a function of centrality for CGC/KLN and
  $\sim N_{\rm part}$ bulk density distributions, respectively. PHENIX
  data~\protect\cite{phenixRaa} correspond to hadron transverse
  momenta $p_\perp>4.5$~GeV.
  Bottom: second harmonic
  coefficient of $R_{AA}(p_\perp,\varphi;b)/R_{AA}(p_\perp;b)$.}
\label{fig:highpT}
\end{center}
\end{figure} 
Fig.~\ref{fig:highpT} displays the angular average $R_{AA}(p_\perp;b)$
of~(\ref{raaunav}) as well as its second harmonic, denoted as
$v_2(p_\perp;b)$. The former is rather insensitive to the shape of the
``overlap'' region as it involves an integral over the azimuth
$\varphi$. The asymmetry $v_2$ for semi-central collisions is slightly
larger for CGC versus Glauber initial conditions. Nevertheless, for
the assumed non-dissipative expansion, which leads to a
rather rapid cooling of the medium, the difference is moderate.

\section{Conclusions}
We have shown that the eccentricity predicted by CGC models is larger
than that from Glauber models.  This property is due to the different
scaling behavior of the density in the transverse plane of the
collision.  Namely, $k_\perp$-factorization predicts that
$dN/d^2r_\perp$ is proportional to the smaller rather than to the
average of the two local saturation scales.  We also find that the
centrality dependence of the multiplicity and of the eccentricity
depend only weakly on ``details'' of the unintegrated gluon
distribution function, such as leading-twist shadowing and the
anomalous dimension. The centrality dependence of the saturation
momentum can be tested further by high-$p_\perp$ particle production
in p+A collisions~\cite{DMTS}.

These observations could be very important for our understanding of
the collective expansion of the high-density matter created at RHIC
and, in the near future, at the LHC.  Larger initial eccentricity
should lead to larger $v_2$ in the final state unless the build-up of
azimuthally asymmetric collective flow is damped by dissipative
effects.  Ref.~\cite{Hirano:2005xf} found that dissipation in the
hadronic stage might not be sufficient to explain experimental
data.  Dissipative corrections to relativistic ideal fluid expansion
of a QGP have recently been discussed in refs.~\cite{BBBO05}.
Previous parton cascade simulations with only elastic two-body
scattering concluded that a huge transport opacity (large parton cross
sections and/or large parton densities) is needed to account for the
measured $v_2$~\cite{molnar01}.  It will be most interesting to
reconsider those results using the CGC initial conditions suggested
here together with parton multiplication
processes~\cite{Xu:2004mz}. 

Finally, we point out that a better understanding of {\em
large}-$x$ effects within the CGC is important to provide quantitative
results on the structure in coordinate space of the initial condition
for hydrodynamic models of heavy-ion collisions at RHIC and LHC.

\vspace*{2cm}{\bf Note added:}
This note has been added in compliance with a request of the referee of this
paper to clarify our use of the participant density~(\ref{eq:npart})
in the definition~(\ref{eq:qs}) of the saturation momentum, and in the
unintegrated gluon distribution function~(\ref{eq:uninteg}). This may
raise concern as the unintegrated gluon distribution functions
$\phi_A$ and $\phi_B$ then appear to be non-universal quantities which
depend on properties of the collision partner. One might prefer to
take $Q_{s\;A,B}^2$ to be proportional to the thickness function
$T_{A,B}$ of the corresponding nucleus rather than to its participant
density. 

We have found that doing so leads to a significant overestimate of the
multiplicity in peripheral collisions and hence to a less good
description of the data shown in Fig.~\ref{fig:dndy} than for the original
KLN approach~\cite{KLN01} with $Q_s^2\sim n_{\rm part}$. This failure
can be corrected by taking into account explicitly the probabilities
$p_{A,B}<1$ of actually encountering at least one nucleon in nucleus
$A$ or $B$, respectively, at a given transverse coordinate. This
redefines the unintegrated gluon distributions as $\Phi_{A,B} =
p_{A,B} \, \phi_{A,B}$. The ${\bf r}_\perp$-dependence of $Q_s^2$ can
then be defined as $T_{A,B}/p_{A,B}$, which prevents it from dropping
below the saturation scale for a single nucleon at the periphery of
the nuclei. Factorization is now manifestly preserved and we also find
that the centrality dependence of the multiplicity is very similar
to that from the original KLN approach, shown in
Fig.~\ref{fig:dndy}.

With respect to the multiplicity $\tilde{N}\equiv dN/d^2{\bf r}_\perp
dy$, however, the approach sketched in the previous paragraph is
essentially identical to that of ref.~\cite{KLN01}. To realize this,
note that the multiplicity is, to a good approximation, a homogeneous
function of first degree: $\tilde{N}(\kappa Q_{s\,A}^2,\kappa
Q_{s\,B}^2) \simeq\kappa^m \,\tilde{N}(Q_{s\,A}^2,Q_{s\,B}^2)$, with
$m=1$. This can be easily understood from the approximate form
$\tilde{N}\sim{\rm min}(Q_{s\,A}^2,Q_{s\,B}^2)\;
|\log\,Q_{s\,A}^2/Q_{s\,B}^2|$, for example. Since now $p_A \, p_B\,
Q_{s\,B}^2$ is proportional to $p_A \, T_B= n^B_{\rm part}$, we recover
$p_A \, p_B\, \tilde{N}(T_A/p_A,T_B/p_B) \simeq \tilde{N}(n^A_{\rm
part},n^B_{\rm part})$.

The transverse energy $dE_\perp/d^2{\bf r}_\perp dy \sim {\rm
min}(Q_{s\,A}^2,Q_{s\,B}^2)\; {\rm max}(Q_{s\,A},Q_{s\,B})\;
|\log\,Q_{s\,A}^2/Q_{s\,B}^2|$, which we use as a weight in
eq.~(\ref{eq:ecc}), is homogeneous to degree $m=3/2$ rather than 1.
Consequently, we find that the approach sketched in this note predicts
an eccentricity roughly half-way inbetween the Glauber and the KLN
approaches for the most peripheral collisions but is close to the
latter for semi-central collisions where $T_{A,B}$ and $n^{A,B}_{\rm
  part}$ differ very little.

\vspace*{1cm}
\begin{acknowledgments}
Y.~N.\ acknowledges support from GSI and DFG.
H.~J.~D.\ is supported by the BMBF-Forschungsvorhaben
05-CU5RI1/3. A.~A.\ thanks S.~Wicks and W.~Horowitz for useful
discussions on energy loss.
\end{acknowledgments}

\clearpage

\begin{figure}[tb]
\begin{center}
\includegraphics[width=9cm,clip]{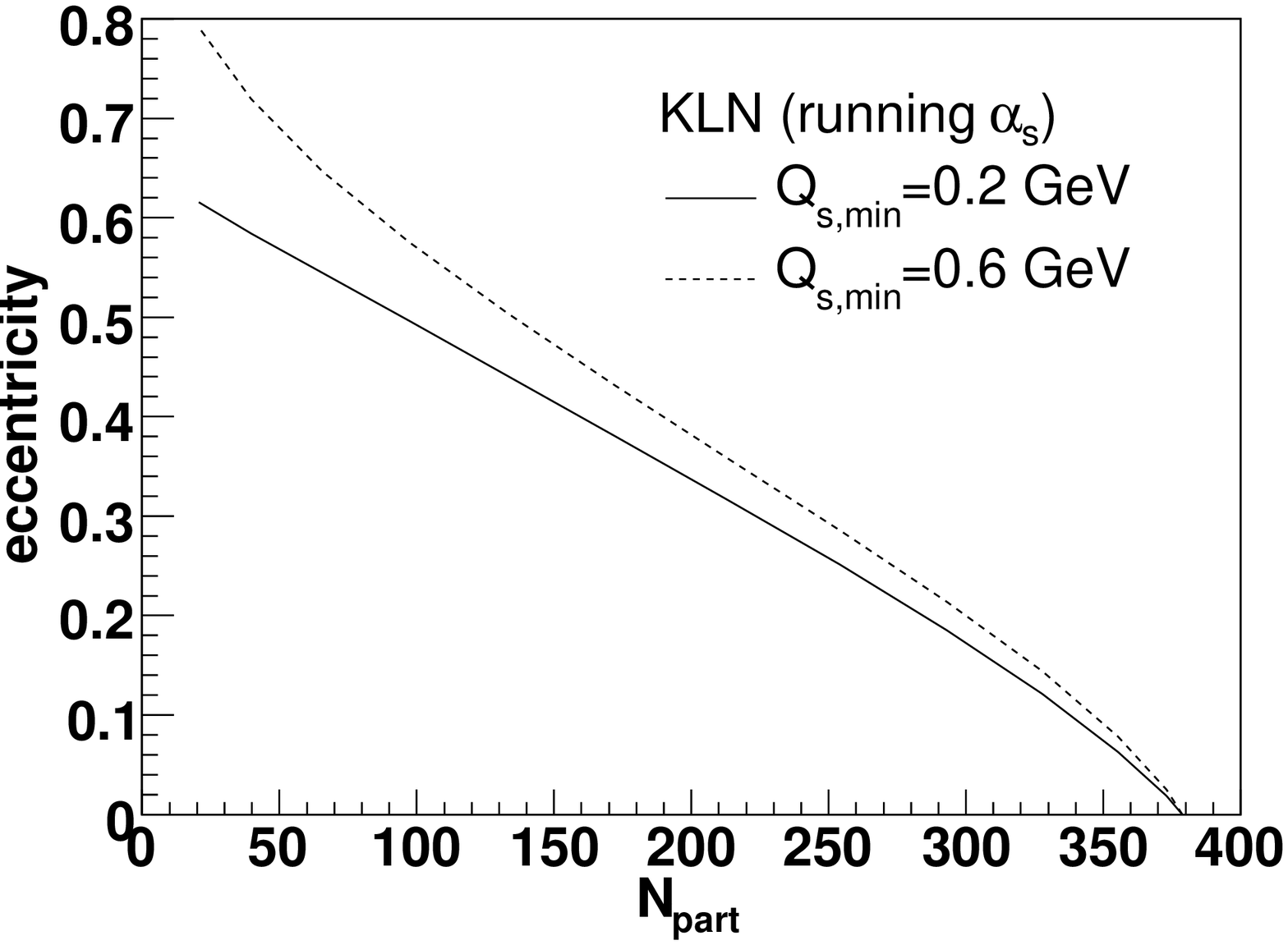}
\caption{{\bf Additional Figure:} Eccentricity as a function of
  $N_{\rm part}$ for the KLN-model of the unintegrated gluon
  distribution function. The cut-off in the integrations over
  $k_\perp$ and $p_\perp$ in eq.~(\ref{eq:ktfac}) is varied from our
  ``standard'' choice of $Q_s(x_i)>\Lambda_{\rm QCD}$ to 
  $Q_s(x_i)>3\Lambda_{\rm QCD}$. Note that a larger cut-off
  effectively chops off part of the low-density rim of the ``overlap''
  region. The eccentricity then increases slightly.
}
\label{fig:QsminCheck}
\end{center}
\end{figure} 

\begin{figure}[tb]
\begin{center}
\includegraphics[width=9cm,clip]{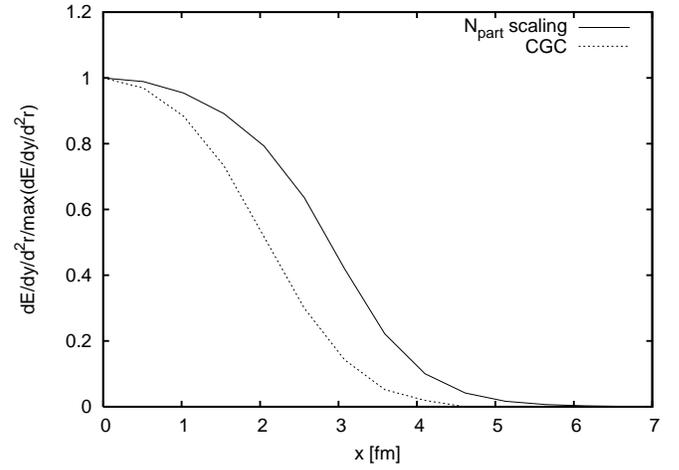}
\caption{{\bf Additional Figure:} The mid-rapidity energy density
  profile for a Au+Au collision at $b=8$~fm as obtained from
  eq.~(\ref{eq:ktfac}) with the KLN unintegrated gluon distribution or
  from a Glauber model, respectively.
  }
\label{fig:edensProfile}
\end{center}
\end{figure} 


\begin{thebibliography}{99}
\bibitem{EtQsmax}
The energy density $dE_\perp/d^2\bm{r}_\perp dy$ in fact
involves another factor of $Q_s^{\rm max}$ which softens the fall-off
along $r_x$ somewhat. However, this is inessential for the
present qualitative discussion.
\bibitem{LargeX} The situation is different for the forward region of
$p+A$ collisions (and $p_\perp\gg\Lambda_{\rm QCD}$) where it is
possible to treat the proton within full DGLAP~\cite{DHJ}, which
respects the momentum sum rule and accounts for $Q^2$-dependent
large-$x$ effects.
\bibitem{commentRaa} Strictly speaking, eq.~(\ref{raaunav}) is valid
only when $n_g=n_q$. For $n_g-n_q\equiv\delta n\neq0$ but small
($\delta n\log p_\perp \ll1$) a correction factor of
$1-(1-f_g)\,\delta n\log p_\perp$ appears on the r.h.s.\ of that
equation. Here, we focus on the sensitivity of $R_{AA}$ to the shape
of the bulk in the transverse plane at fixed $p_\perp$; we therefore
absorb this factor into the overall normalization, which is set to
reproduce $R_{AA}\approx0.2$ for the most central collisions.
\end{thebibliography}
\end{document}